\documentclass{ws-ijmpa}

\begin{document}

\markboth{B. Krusche}
{In-medium Properties of Hadrons}

%
\catchline{}{}{}{}{}
%

\title{IN-MEDIUM PROPERTIES OF HADRONS}

\author{B. Krusche}

\address{Department of Physics and Astronomy, University of Basel,
Klingelbergstr. 82\\
CH-4056 Basel, Switzerland\\
Bernd.Krusche@unibas.ch}

\maketitle

\begin{history}
\received{Day Month Year}
\revised{Day Month Year}
\end{history}

\begin{abstract}
A diverse experimental program for the study of
the photoproduction of mesons off nuclei has been carried out - and is still
running - at the Mainz MAMI and Bonn ELSA electron accelerators with the 
TAPS, Crystal Barrel, and Crystal Ball calorimeters. It is motivated
as a detailed study of the in-medium properties of hadrons and the 
meson - nucleus interactions. Typical examples
for the in-medium behavior of vector mesons ($\omega$), scalar mesons
($\sigma$), and nucleon resonances (P$_{33}$(1232), S$_{11}$(1535), 
D$_{15}$(1520)) are discussed. Special attention is paid to meson - nucleus
final state interactions. 
\end{abstract}

\ccode{PACS numbers: 13.60.Le, 25.20.Lj, 21.65+f}

\section{Introduction}	

In-medium properties of hadrons are a hotly debated topic since they
are closely related to the properties of low-energy, non-perturbative QCD. 
Unlike any other composite systems, hadrons are objects, which are build out 
of constituents with masses (5 - 15 MeV for u,d quarks) that are negligible
compared to the total mass. Most of the mass is generated by dynamical effects 
from the interaction of the quarks and an important role is played by the
spontaneous breaking of chiral symmetry, the fundamental symmetry of QCD.
The symmetry breaking, which is connected to a non-zero expectation value 
of scalar $q\bar{q}$ pairs in the vacuum, the chiral condensate, is  
reflected in the hadron spectrum. Without it, hadrons would appear as 
mass degenerate parity doublets, which is neither true for baryons nor for 
mesons. However, model calculations (see e.g. Ref~\refcite{Lutz_92}) indicate 
a temperature and density dependence of the condensate, which is connected with 
a partial restoration of chiral symmetry. Although there is no direct relation 
between the quark condensate and the in-medium properties of hadrons 
(masses, widths etc.), there is an indirect one via QCD sum rules, which connect 
the QCD picture with the hadron picture. In the latter, the in-medium 
modifications arise from the coupling of mesons to resonance - hole states 
and the coupling of the modified mesons to resonances. The best known example 
is the treatment of the $\Delta$ in the framework of the $\Delta$-hole model 
(see e.g. Ref.~\refcite{Oset_83,Koch_83}). Recently, Post, Leupold, and 
Mosel\cite{Post_04} have calculated the hadron in-medium spectral 
functions for $\pi$-, $\eta$-, and $\rho$-mesons and baryon resonances in a 
self-consistent coupled channel approach.
\begin{figure}[hbt]
\centerline{\psfig{file=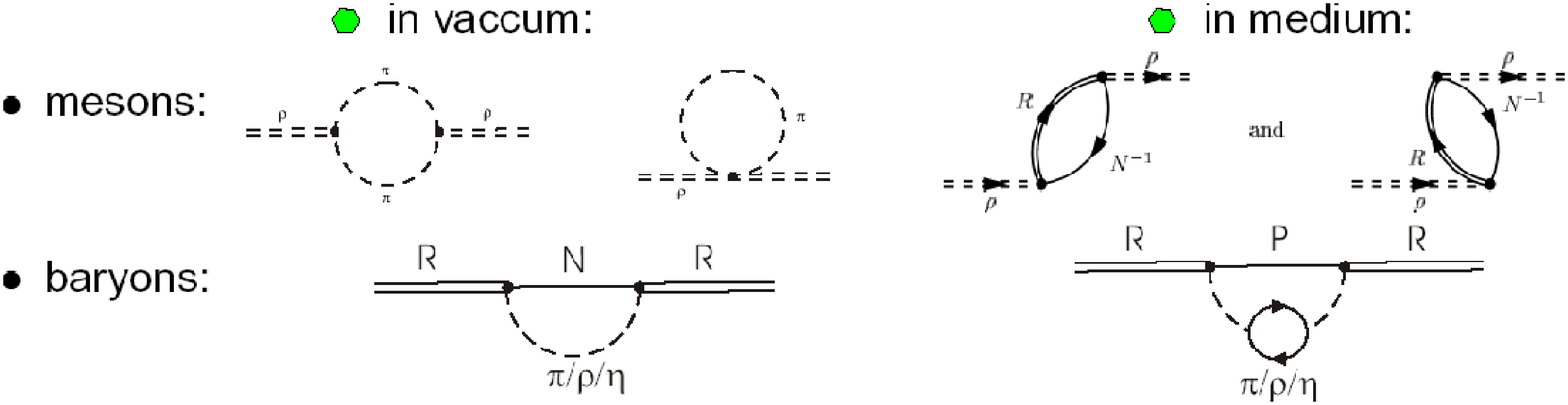,width=10.cm}}
\centerline{\psfig{file=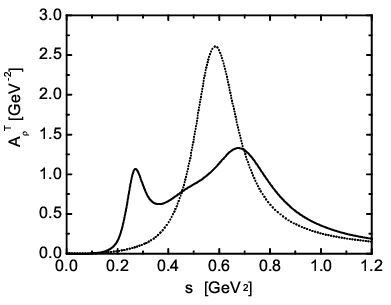,height=3.5cm}
\psfig{file=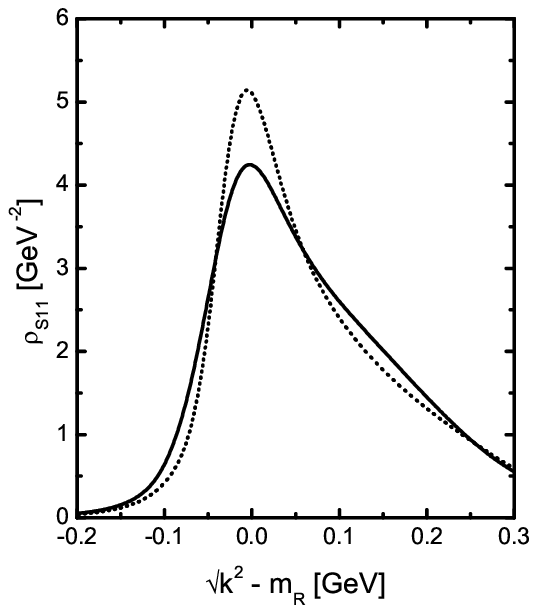,height=3.4cm}
\psfig{file=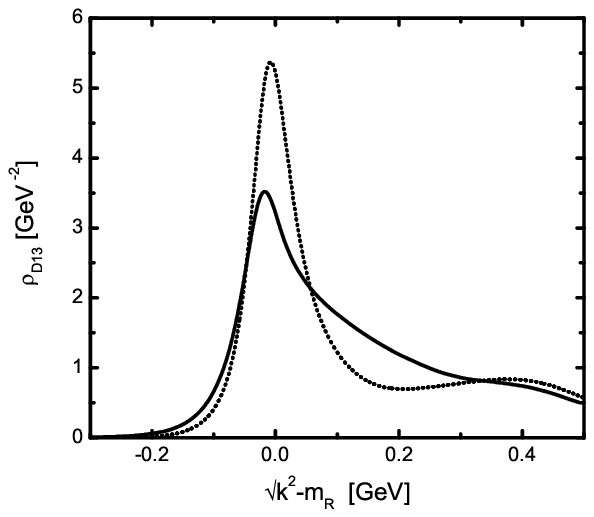,height=3.4cm}}
\vspace*{8pt}
\caption{Upper part: diagrams for the vacuum and in-medium self-energies
of mesons and baryons.\protect\cite{Post_04} Lower part: predicted vacuum 
(dashed curves) and in-medium (solid curves) spectral functions for the 
$\rho$-meson (left), the S$_{11}$(1535) (center), and the D$_{13}$(1520) 
(right) resonances at $q$=0.\protect\cite{Post_04}
\label{fig:selfen}}
\end{figure}
The relevant diagrams for the vacuum and in-medium self-energies and typical
results for the spectral functions are shown in Fig.~\ref{fig:selfen}. In case 
of the resonances, the largest effects are predicted for the D$_{13}$, due 
to the strong coupling of this resonance to the $N\rho$ channel.

\begin{figure}[thb]
\begin{minipage}{4cm}
\centerline{\psfig{file=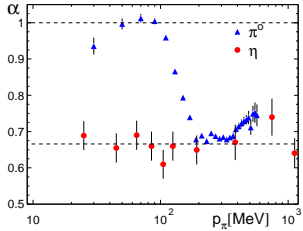,height=3.cm}}
\end{minipage}
\hspace*{1cm}\begin{minipage}{6cm}
\caption{Scaling coefficient $\alpha$ derived from a fit of 
$\sigma (A)\propto A^{\alpha}$ to data from carbon, calcium, niobium and lead
nuclei, as function of meson momentum for $\pi^o$- and 
$\eta$-mesons.\protect\cite{Mertens_06}\protect\cdash\protect\cite{Krusche_04}
\label{fig:alpha}}
\end{minipage}
\end{figure}

The experimental investigation of hadron in-medium properties is  
complicated by initial (ISI) and/or final (FSI) state interactions. Since the 
experiments discussed here use photoproduction of mesons, no ISI but significant 
FSI effects must be considered. However, the investigation of this reactions 
allows also a detailed study of the meson - nucleus interactions, which are 
responsible for the FSI.\cite{Roebig_96}\cdash\cite{Krusche_04a}
The simplest information comes from the mass number scaling of the meson 
production cross sections, which can be parameterized by 
$\sigma (A)\propto A^{\alpha}$. Qualitatively, $\alpha\approx$2/3 corresponds 
to strong absorption (cross section proportional to nuclear surface) and 
$\alpha\approx$1 to negligible absorption (cross section proportional to nuclear 
volume). The scaling for different meson production channels has been 
discussed in detail in Refs.~\refcite{Krusche_04,Krusche_04a}. Fig. \ref{fig:alpha}
shows the coefficient $\alpha$ for $\pi^o$- and $\eta$ mesons as 
function of their momentum. The absorption of pions is strong in the 
range where the $\Delta$ resonance can be excited (around 200 MeV), but nuclear 
matter becomes almost transparent for pions below momenta of 100 MeV. The 
absorption of $\eta$ mesons is already large for very low momenta, since the 
S$_{11}$ resonance is located at the $\eta$ production threshold.
 Quantitatively,
one can convert the coefficients into absorption cross sections
and mean free paths e.g. via Glauber theory (see e.g. Ref.~\refcite{Roebig_96}). 

An extensive program for the investigation of meson
photoproduction reactions off nuclei has been carried out at the Mainz MAMI and
Bonn ELSA accelerators with the TAPS\cite{Novotny_91,Gabler_94} and 
Crystal Barrel detectors\cite{Aker_92}. It included the following topics: 
\begin{itemize}
\item{The search for in-medium modifications of vector mesons
via the investigation of the resonance shape of the $\omega$ meson
from its $\pi^o\gamma$ decay.\cite{Trnka_05}}
\item{The study of the pion-pion invariant mass distributions for
$2\pi^o$ and $\pi^o\pi^{\pm}$ photoproduction as a tool for  
the in-medium behavior of the $\sigma$ meson.\cite{Messchendorp_02,Bloch_06}}
\item{Resonance contributions to $\eta$, $\pi$, $2\pi$
meson production reactions from nuclei, aiming at the in-medium properties of
nucleon 
resonances.\cite{Mertens_06}\cdash\cite{Krusche_04a,Krusche_05,Krusche_01}
}
\item{The search for $\eta$-nucleus bound states 
($\eta$ mesic nuclei).\cite{Hejny_01,Pfeiffer_04}}
\item{The investigation of $\eta$, $2\pi^o$, and $\eta'$ photoproduction off 
light nuclei for the study of the excitation of resonances of the 
neutron.\cite{Krusche_95a}\cdash\cite{Weiss_03}
}
\end{itemize}

\section{In-Medium Modifications of Mesons}

\subsection{Vector mesons: the $\omega$-meson}

In-medium modifications of vector mesons have been searched for via the
spectroscopy of di-lepton pairs in heavy ion reactions by the CERES
experiment\cite{Agakichiev_95,Adamova_03} and more recently the NA60 
collaboration\cite{Damjanovic_05}, which reported an in-medium broadening of the
$\rho$. In-medium modifications of $\rho$ and $\omega$ mesons were reported 
from 12 GeV p+A reactions at KEK\cite{Naruki_06}. Recently, a modification of 
the $\Phi$ meson has been suggested on the basis of the $A$-dependence of the 
photoproduction yields.\cite{Ishikawa_05}

\begin{figure}[tbt]
\centerline{\psfig{file=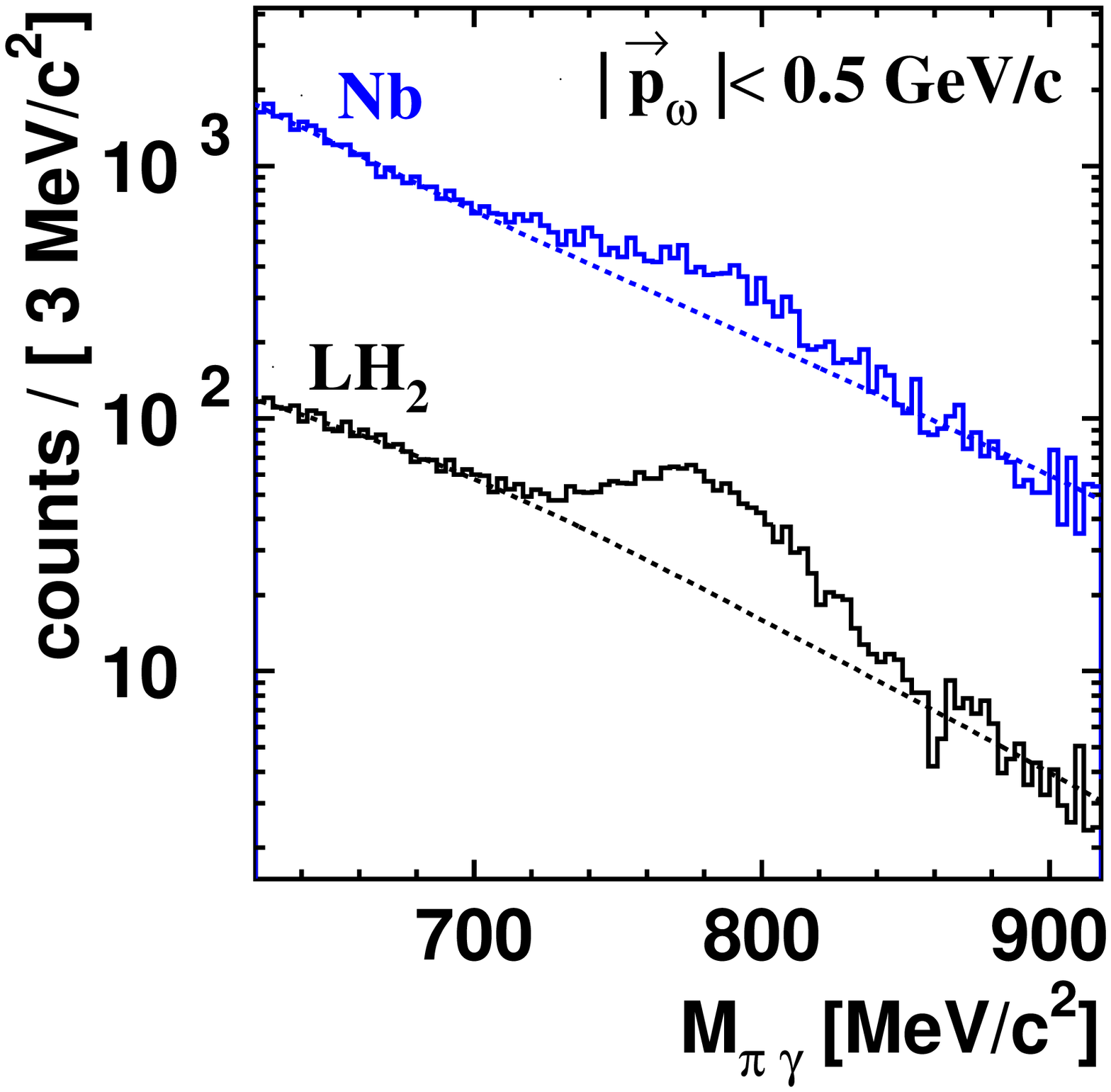,height=4.cm}
\psfig{file=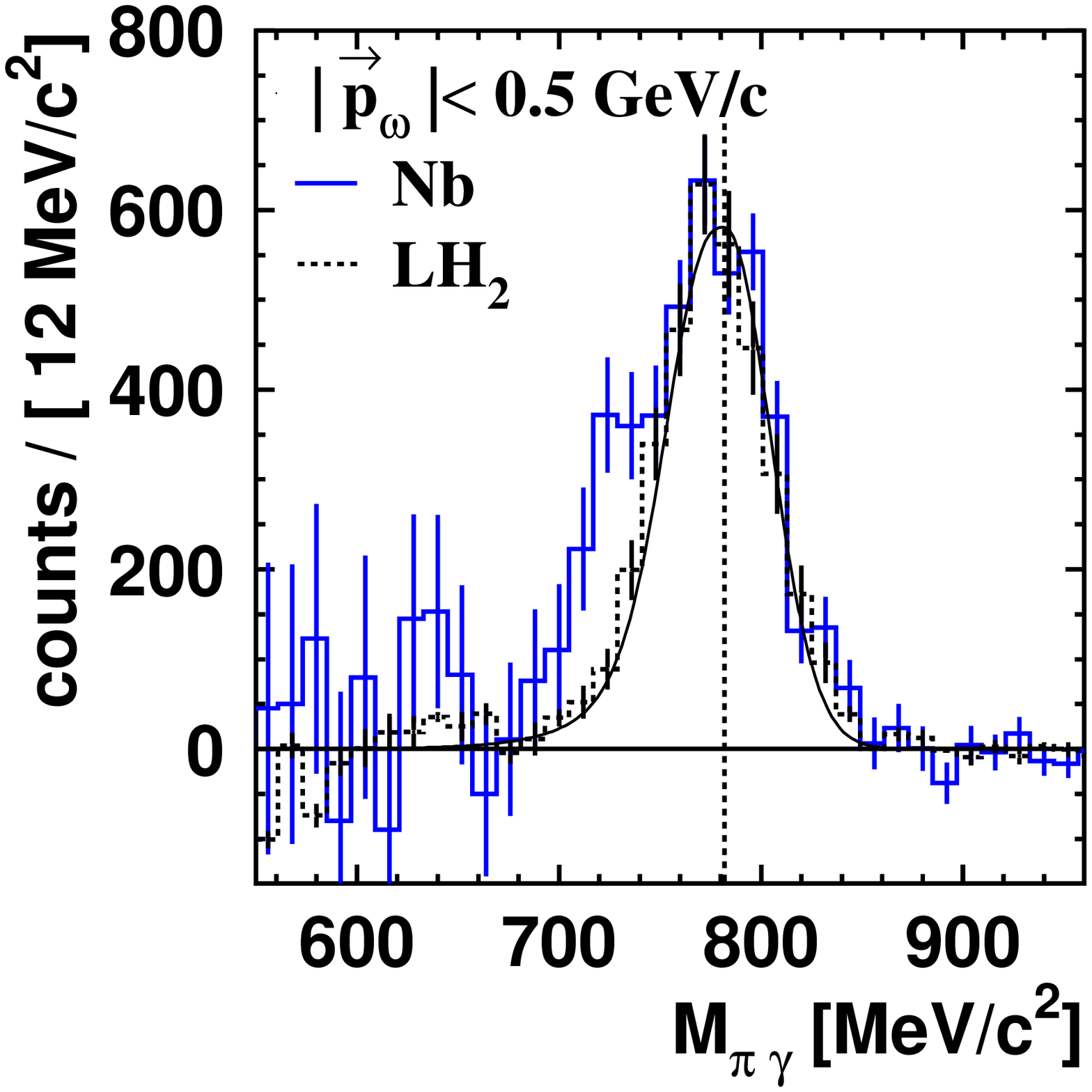,height=4.cm}
\psfig{file=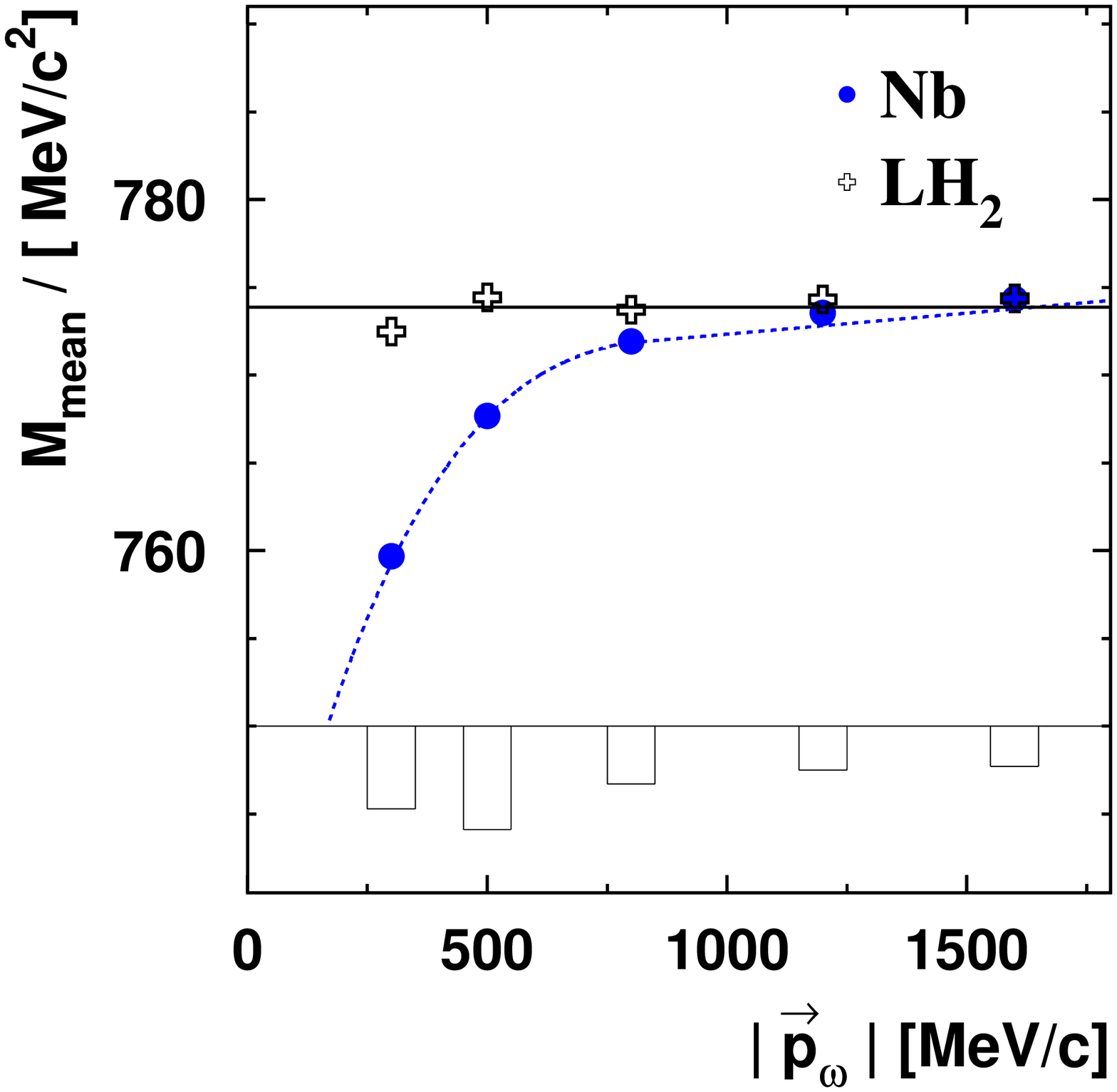,height=3.7cm}}
\vspace*{8pt}
\caption{$\omega$ invariant mass peak. Left: comparison 
invariant mass spectra niobium - proton, center: after subtraction 
of (combinatorial) background, right: momentum dependence of peak 
centroid.\protect\cite{Trnka_05}  
\label{fig:omega}}
\end{figure}

The CBELSA/TAPS\cite{Trnka_05} experiment measured the line shape of the 
$\omega\rightarrow\pi^o\gamma$ invariant mass peak off the free proton and off 
nuclei. The measured invariant mass peaks of the 
$\pi^o\rightarrow\gamma\gamma$, $\eta\rightarrow 3\pi^o\rightarrow 6\gamma$, and 
$\eta'\rightarrow\pi^o\pi^o\eta\rightarrow 6\gamma$ decays were identical
for hydrogen and the nuclear targets. Only in case of the $\omega$ 
a low energy shoulder of the peak was found for nuclei (see Fig.
\ref{fig:omega}). Due to its life time, only a small fraction of 
the $\omega$'s decay in the medium, so that also the nuclear invariant mass 
peaks include a dominant contribution of unmodified in-vacuum decays. 
This is reflected in the data since the low-energy tail of the invariant mass 
stems almost exclusively from $\omega$'s with small momenta, which have the 
largest change for in-medium decays. The peak centroids as function of the 
momentum are shown in Fig. \ref{fig:omega}, right hand side. Similar results 
where found for carbon, an analysis of the $A$ scaling of the production 
cross sections is under way, and a second generation experiment is approved.  

\subsection{Scalar mesons: the $\sigma$-meson}

\begin{figure}[thb]
\centerline{\psfig{file=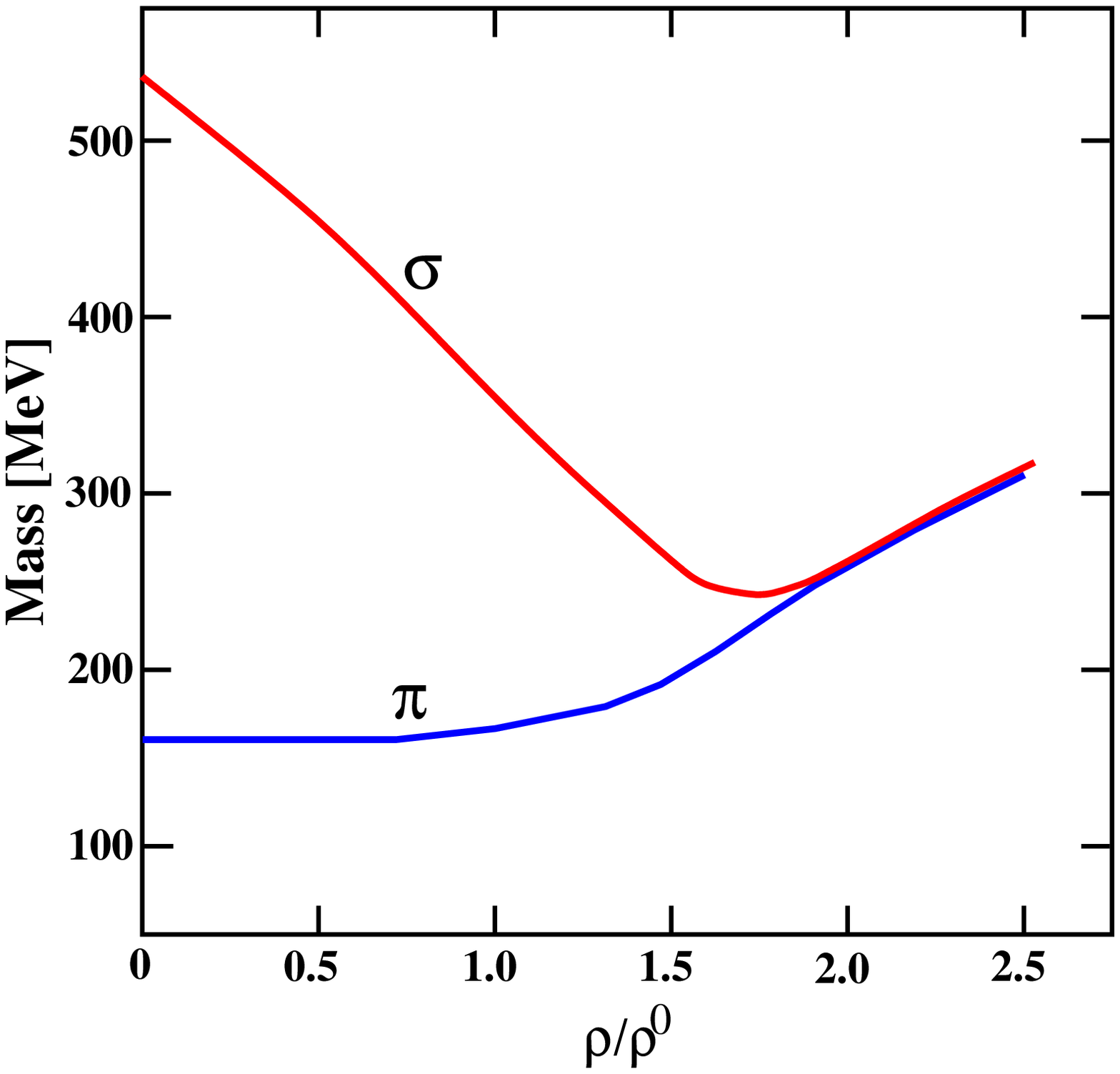,height=3.5cm}
\psfig{file=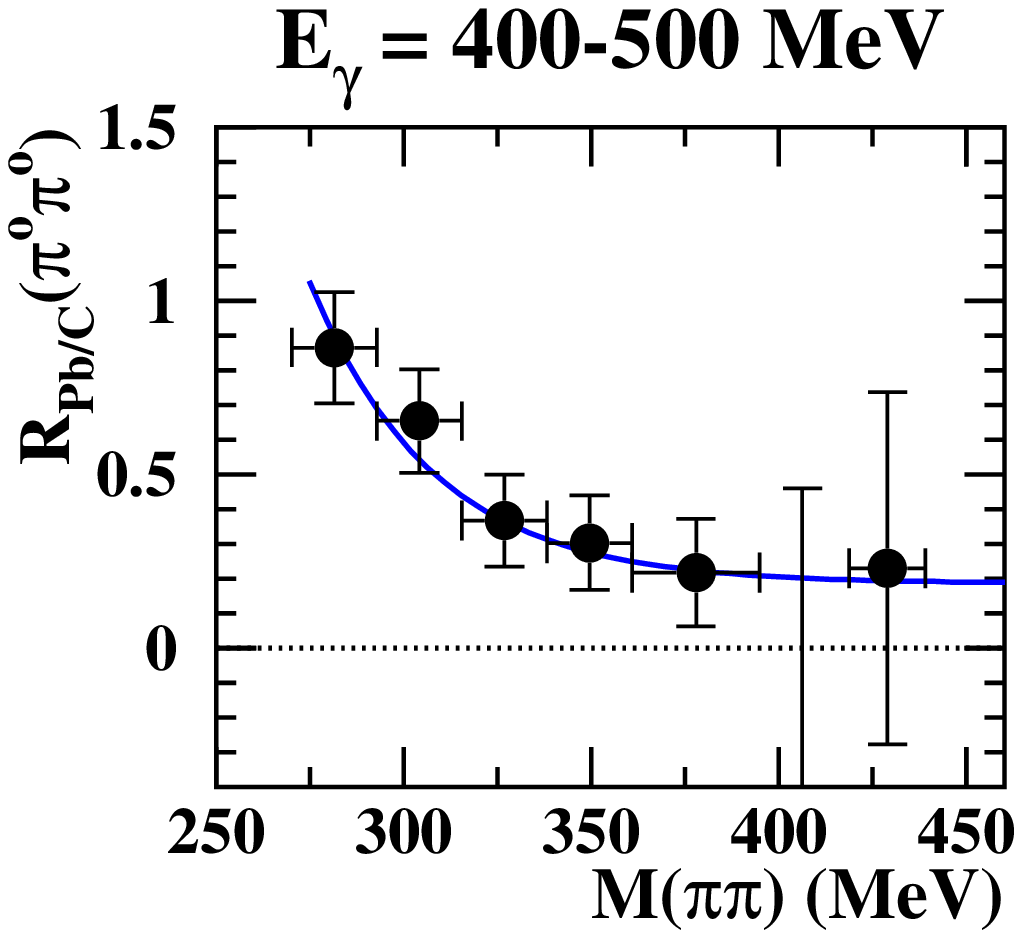,height=4.cm}
\psfig{file=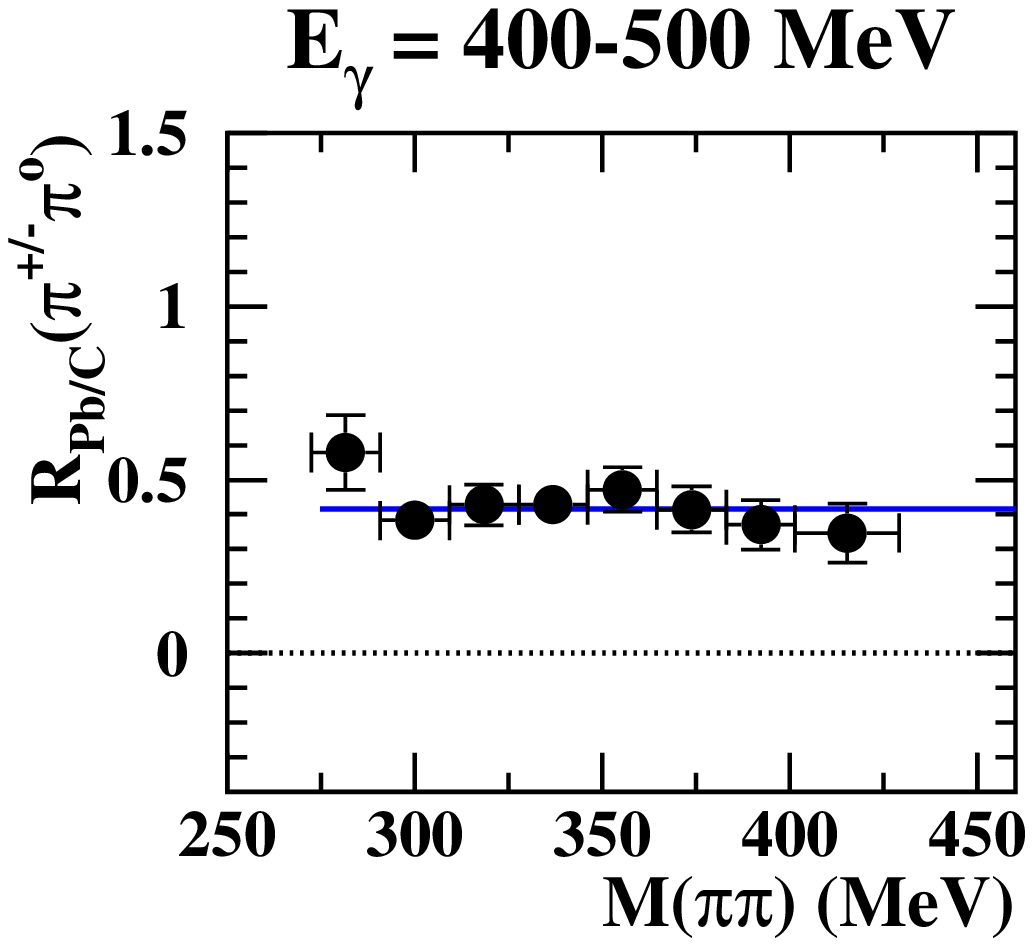,height=4.cm}}
\vspace*{8pt}
\caption{Left hand side: predicted density dependence of $\pi$ and $\sigma$
mass\protect\cite{Bernard_87}, center (right hand side): measured $R_{Pb/C}$
ratio for $\pi^o\pi^o$ ($\pi^{\pm}\pi^o$).\protect\cite{Messchendorp_02} 
\label{fig:2pold}}
\end{figure}

A very interesting case is the mass-split between the $J^{\pi}$=0$^{-}$ 
pion and the $J^{\pi}$=0$^{+}$ $\sigma$-meson. The naive assumption, that the 
two masses should become degenerate in the chiral limit is supported by model
calculations. A typical result is the density dependence of the mass calculated
in the Nambu-Jona-Lasino model\cite{Bernard_87} (see Fig. \ref{fig:2pold}). 
The nature of the $\sigma$ meson itself has been much discussed in 
the literature. The review of particle properties\cite{Eidelmann_04} lists as 
$\sigma$ the $f_{0}$(600) with a mass of 400 - 1200 MeV and a full width 
between 600 MeV - 1000 MeV. Precise predictions for mass and width from 
dispersion relations were derived in Ref.~\refcite{Caprini_06}. In some 
approaches the $\sigma$ is treated as a pure $q\bar{q}$ (quasi)bound 
state\cite{Bernard_87,Hatsuda_99,Aouissat_00}, in others as a correlated 
$\pi\pi$ 
pair in a $I=0$, $J^{\pi}=0^+$ state\cite{Chiang_98}\cdash\cite{Chanfray_06}. 
But in any case a strong coupling to scalar, iso-scalar pion pairs is predicted. 
As a consequence, all models predict a downward shift of the strength in the 
invariant mass distributions of such pion pairs in the medium. This is either 
due to the in-medium spectral function of the $\sigma$ meson\cite{Hatsuda_99} 
or the modification of the pion-pion interaction\cite{Roca_02} due to the 
coupling to nucleon - hole, $\Delta$ - hole and $N^{\star}$ - hole states. 

First experimental evidence has been reported by the CHAOS collaboration from 
pion induced double pion production.\cite{Bonutti_96}\cdash\cite{Grion_05} 
They found a build-up of strength with rising mass number at low
invariant masses for the $\pi^+\pi^-$ final state, but not for the 
$\pi^+\pi^+$ channel where the neutral $\sigma$ meson cannot contribute. 
A similar effect was found in the $\pi^- A\rightarrow A\pi^0\pi^0$ reaction
with Crystal Ball at BNL.\cite{Starostin_00} 

\begin{figure}[tbt]
\centerline{\psfig{file=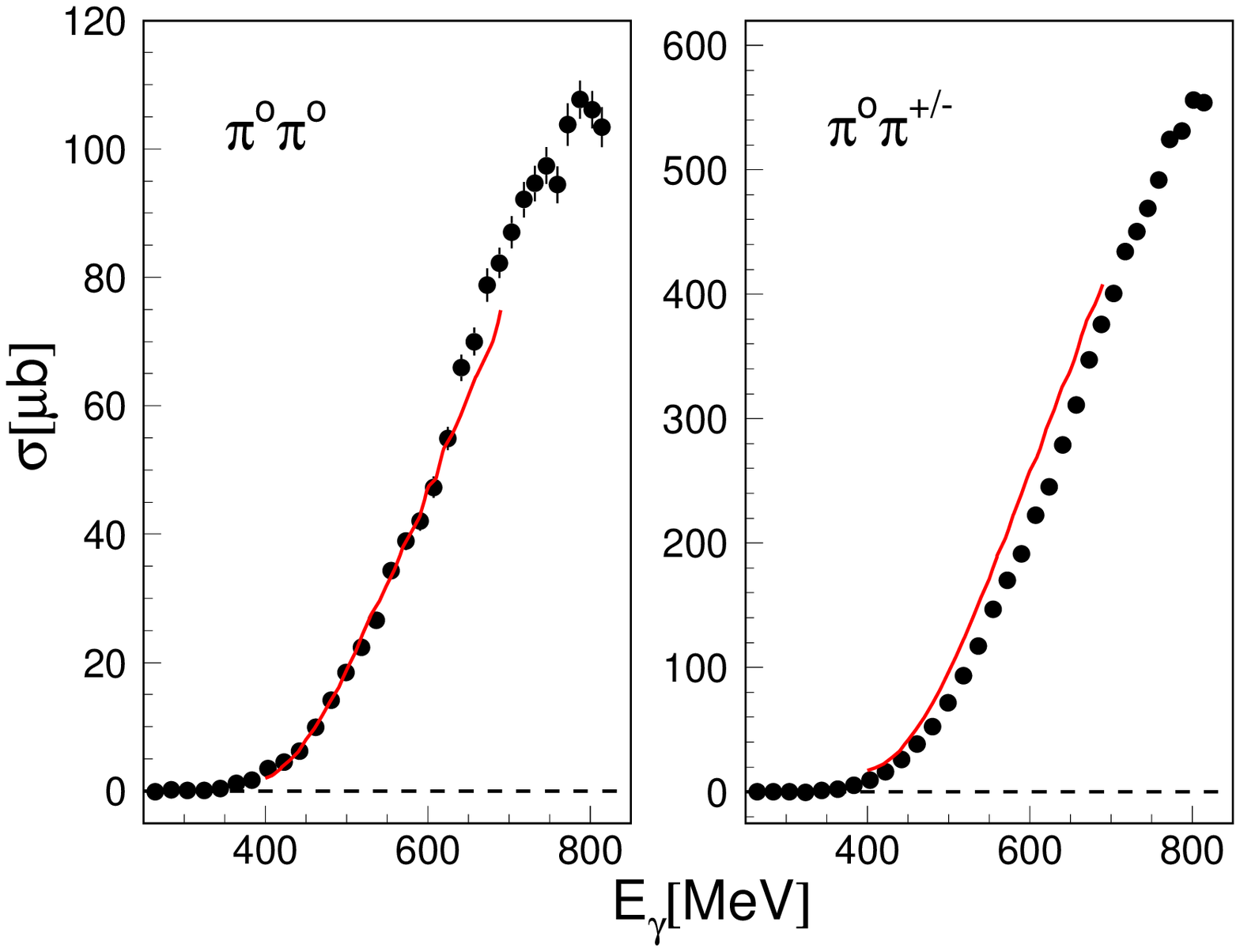,height=4.5cm}
\psfig{file=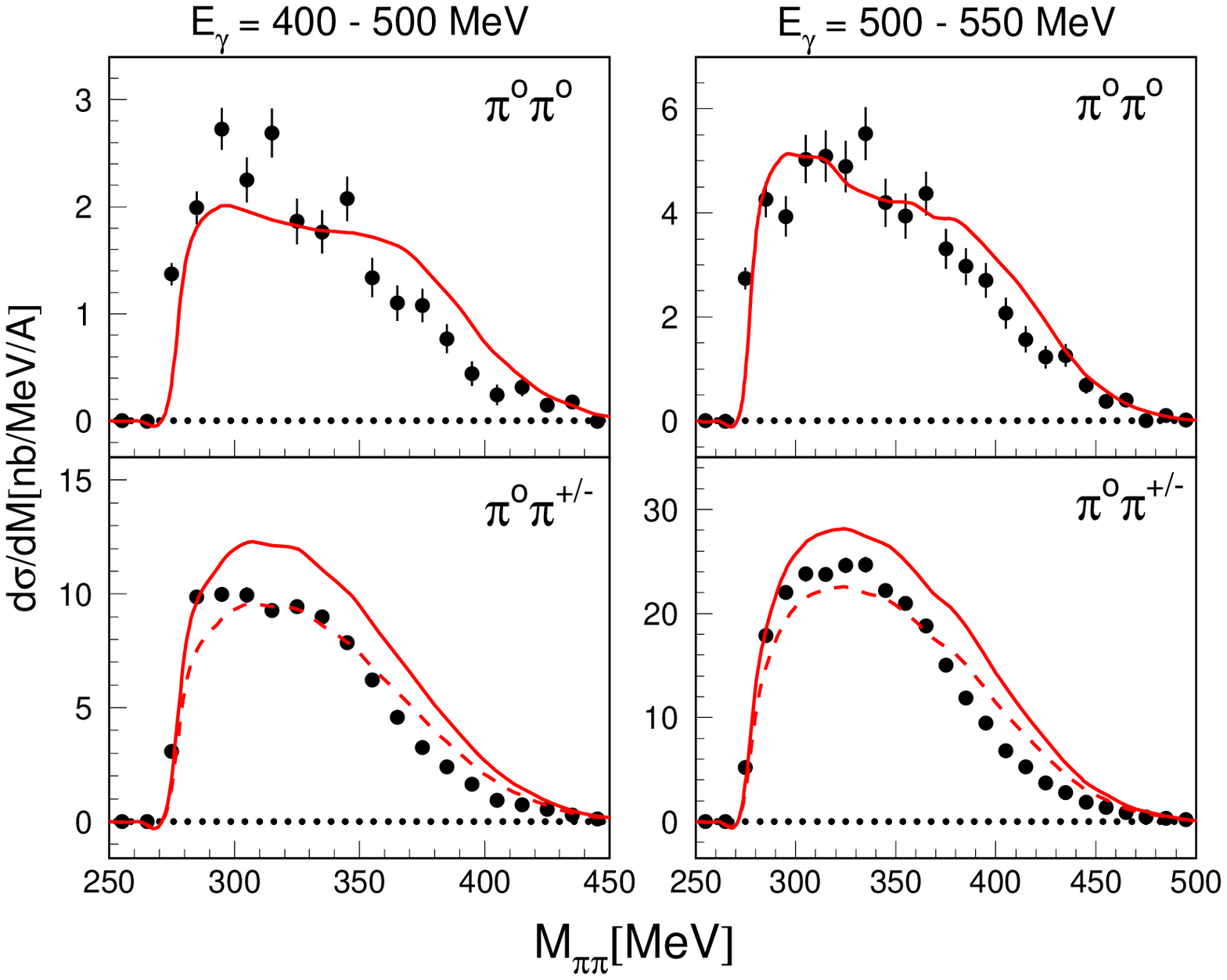,height=4.5cm}}
\vspace*{8pt}
\caption{Double pion production off $^{40}$Ca. Left hand side: total cross 
sections, right hand side: invariant mass distributions. Data from
Ref.~\protect\refcite{Bloch_06}, curves: BUU results\protect\cite{Buss_06} 
(dashed: re-scaled to data).
\label{fig:2ptotal}}
\end{figure}

In contrast to pion induced reactions, photoproduction reactions avoid 
initial state interactions, although also in this case FSI complicates the 
picture. FSI can be minimized by the use of low incident photon energies, 
giving rise to low-energy pions which have large mean free 
paths\cite{Krusche_04} (see Fig. \ref{fig:alpha}).
Photoproduction of pion pairs at low incident photon energies off the free 
proton and the quasifree neutron has been previously studied with the 
DAPHNE\cite{Braghieri_95}\cdash\cite{Ahrens_05}
and TAPS\cite{Krusche_99,Kleber_00,Haerter_97}\cdash\cite{Kotulla_04} detectors
at MAMI in Mainz (see Ref.~\refcite{Krusche_03} for an overview). 
First results from a measurement of double $\pi^0$ and $\pi^0\pi^{\pm}$
photoproduction off carbon and lead have been reported in Ref.
\refcite{Messchendorp_02}. For the heavier nucleus a shift of the strength to 
lower invariant masses was found for the $\pi^0\pi^0$ channel but not for the 
mixed charge channel. This is particularly clearly seen in the normalized ratios
of the invariant mass distributions defined by
\begin{equation}
R_{(Pb/C)}(\pi^o\pi^o) =   
\frac{d\sigma_{Pb}(\pi^0\pi^0)}{\sigma_{Pb}(\pi^0\pi^0) dM} \left/ 
\frac{d\sigma_{C}(\pi^0\pi^0)}{\sigma_{C}(\pi^0\pi^0) dM}
\right.
\end{equation}
and analogously for $R_{(Pb/C)}(\pi^o\pi^{\pm})$. The results are shown in Fig.
\ref{fig:2pold}, center and right hand side. The ratio rises for 
the double $\pi^o$ production, but is almost constant for the mixed charge 
channel. This behavior was used as an argument, that the effect does not arise 
from FSI, which was assumed to be similar for neutral and charged pions. 
In the meantime, also data for the medium-weight nucleus $^{40}$Ca became 
available\cite{Bloch_06} (see Fig. \ref{fig:2ptotal}), which is discussed below.

A quantitative interpretation of the distributions requires models which
account for the `trivial' in-medium effects, like the momentum distributions 
of the bound nucleons and FSI of the pions. 
Recently M\"uhlich et al.\cite{Muehlich_03,Muehlich_04} and Buss 
et al.\cite{Buss_06} have performed detailed calculations of the nuclear 
double pion photoproduction reactions in the framework of coupled channel 
transport models based on the semi-classical Boltzmann-Uehling-Uhlenbeck (BUU) 
equation. Special emphasis was laid on the description of the scattering and 
absorption properties of low energy pions. 
An important result of this model is, that even without explicit in-medium 
modification of the $\pi^o\pi^o$ channel, the respective invariant mass 
distributions show a softening for heavy nuclei. This effect arises solely 
from FSI of the pions. The $\pi N$ absorption cross section increases with pion 
kinetic energy, so that pions with large momenta are more strongly depleted via 
the $\pi N\rightarrow\Delta$, $\Delta N\rightarrow NN$ reaction path. But this 
causes only part of the effect. Important are also 
re-scattering processes, which tend to decrease the pion kinetic energy and thus 
the average pion - pion invariant mass. This effect is enhanced due to charge 
exchange scattering. Since the total cross section for $\pi^o\pi^{\pm}$ production
is much larger than the $\pi^o\pi^o$ cross section, the latter receives 
significant side feeding from the mixed charge channel via 
$\pi^{\pm}N\rightarrow N\pi^o$ scattering. Consequently, the relative softening 
of the $\pi^o\pi^o$ invariant mass by itself is no safe evidence for an 
in-medium modification of the pion - pion interaction. 

The overall agreement between the model results and the measured total cross
sections (see fig. \ref{fig:2ptotal}) is excellent for the $\pi^o\pi^o$ channel. 
The data for the mixed charge channel are a slightly overestimated by the model, 
however, this is within the systematic uncertainty.
The general tendency of a softening of the invariant mass distributions is 
reproduced by the calculations as a consequence of FSI effects.
On a quantitative level, the data for the $\pi^o\pi^o$ invariant masses for the 
lowest incident photon energies seem to be somewhat stronger downward shifted 
than predicted. This is the energy range were the smallest FSI effects are 
expected. 

The search for the possible in-medium modification of the pion - pion
interaction, respectively the $\sigma$ in-medium modification, thus becomes a
quantitative question. At least part of the observed effect is due to 
pion final state interaction. Further efforts are needed to clarify, if there 
are effects beyond what can be explained by BUU.
To this end, data with much higher statistical quality, in particular for lead,
have been recently measured with the Crystal Ball/TAPS detector at MAMI. 

\section{In-Medium Modifications of Nucleon Resonances}

Nucleon resonances in the medium are effected by the Pauli-blocking of final states,
which decreases their width, additional decay channels like $RN\rightarrow NN$
($R$ = resonance, $N$ = nucleon), which cause collisional
broadening, and as already discussed in the introduction, by the coupling to mesons
with medium-modified properties.

\subsection{The $\Delta$ resonance}
The excitation of the $\Delta$ and its propagation through the nuclear medium 
have been studied in different reactions, an in-medium broadening at normal 
nuclear matter density of $\approx$100 MeV has been extracted. In this sense
the $\Delta$ is a well understood test case. However, a detailed understanding of
the $\Delta$ case is also important for the interpretation of meson production
reactions in the higher nucleon resonance regions, where the $\Delta$ contributes 
indirectly e.g. via multiple pion production processes and through re-absorption
of pions.  
\begin{figure}[tbt]
\centerline{\psfig{file=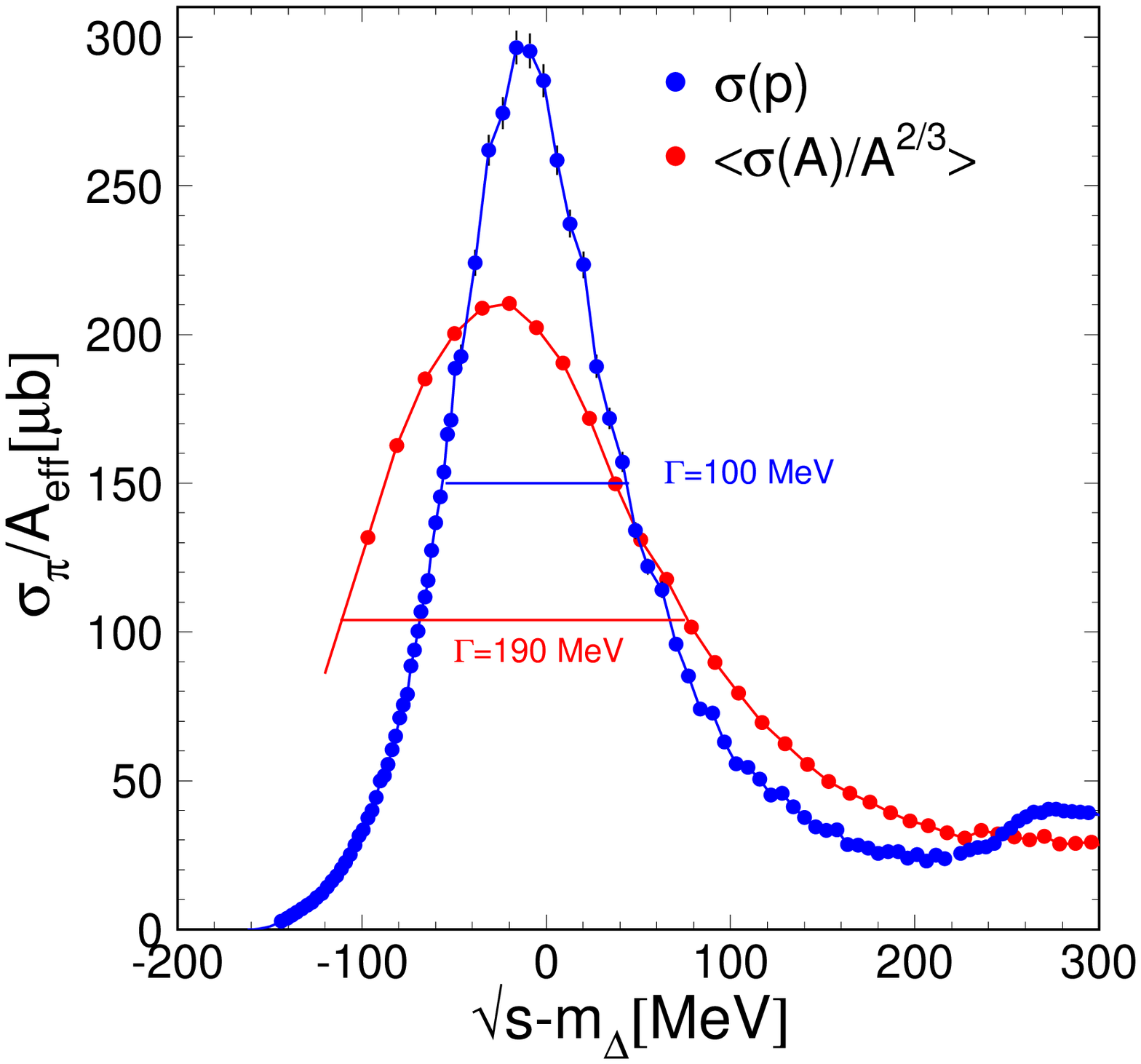,height=3.5cm}
\psfig{file=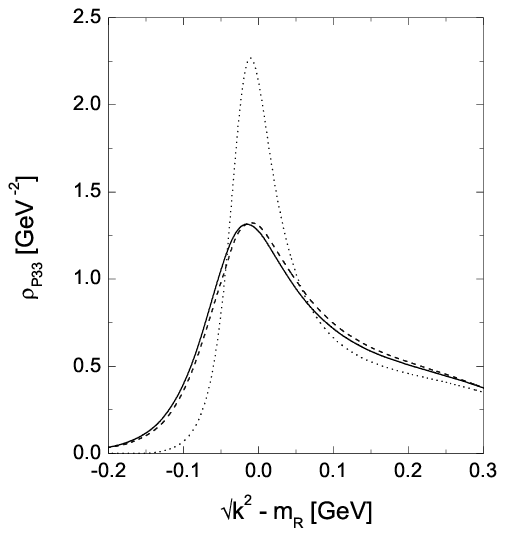,height=3.6cm}}
\vspace*{8pt}
\caption{Left hand side: total $\pi^o$ photoproduction of the proton and the
average for heavy nuclei\protect\cite{Krusche_05}.
 Right hand side: self-consistent spectral functions
in-vacuum and in-medium for the $\Delta$(1232)\protect\cite{Post_04}.
\label{fig:delta}}
\end{figure}
The line shape of the $\Delta$ measured with photoproduction of $\pi^o$ mesons
off the free proton and off nuclei\cite{Krusche_04,Krusche_05} is compared in
Fig. \ref{fig:delta} to the spectral functions calculated in 
Ref.~\refcite{Post_04}. Both, data and model show an in-medium broadening
of the $\Delta$ close to 100 MeV. This is also in agreement with a study
of coherent $\pi^o$ photoproduction of heavy nuclei, which allowed a more 
detailed investigation of the $\Delta$ in-medium properties\cite{Krusche_02}.
 
\subsection{The second resonance region}

The second resonance region is composed of the P$_{11}$(1440), D$_{13}$(1520), 
and S$_{11}$(1535) states. A significant in-medium modification of the D$_{13}$ 
is predicted due to the strong coupling to $N\rho$, while only small effects are 
expected for the S$_{11}$  (see Fig.~\ref{fig:selfen})\cite{Post_04}. The 
experimental finding from total photoabsorption is, that the second resonance 
bump is completely suppressed for all nuclei from beryllium 
onwards.\cite{Bianchi_94} However, 
total photoabsorption does not give any information about the behavior of
individual resonances. Photoproduction of $\eta$ mesons in this energy range
is completely dominated by the S$_{11}$ 
resonance\cite{Krusche_95}\cdash\cite{Crede_05},
while single $\pi^o$, double $\pi^o$ and $\pi^o\pi^{\pm}$
photoproduction\cite{Haerter_97}\cdash\cite{Langgaertner_01,Bartholomy_05}
show a clear signal for the D$_{13}$ resonance. These reactions are therefore 
well suited for the study of the in-medium properties of this resonances. 

Photoproduction of $\eta$ mesons off nuclei had been previously studied up to
800 MeV incident photon energy with TAPS at MAMI \cite{Roebig_96} and for 
energies up to 1.1 GeV at KEK\cite{Yorita_00} and Tohoku\cite{Kinoshita_06}. 
The first experiment found no in-medium broadening of the resonance (beyond 
effects from Fermi smearing and $\eta$ FSI), the KEK experiment reported 
some collisional broadening of the resonance and the Tohoku experiment
pointed to a significant contribution of a higher lying resonance to 
the $\gamma n\rightarrow n\eta$ reaction. However, none of these experiments 
covered the full line shape of the S$_{11}$.

\begin{figure}[tbt]
\centerline{\psfig{file=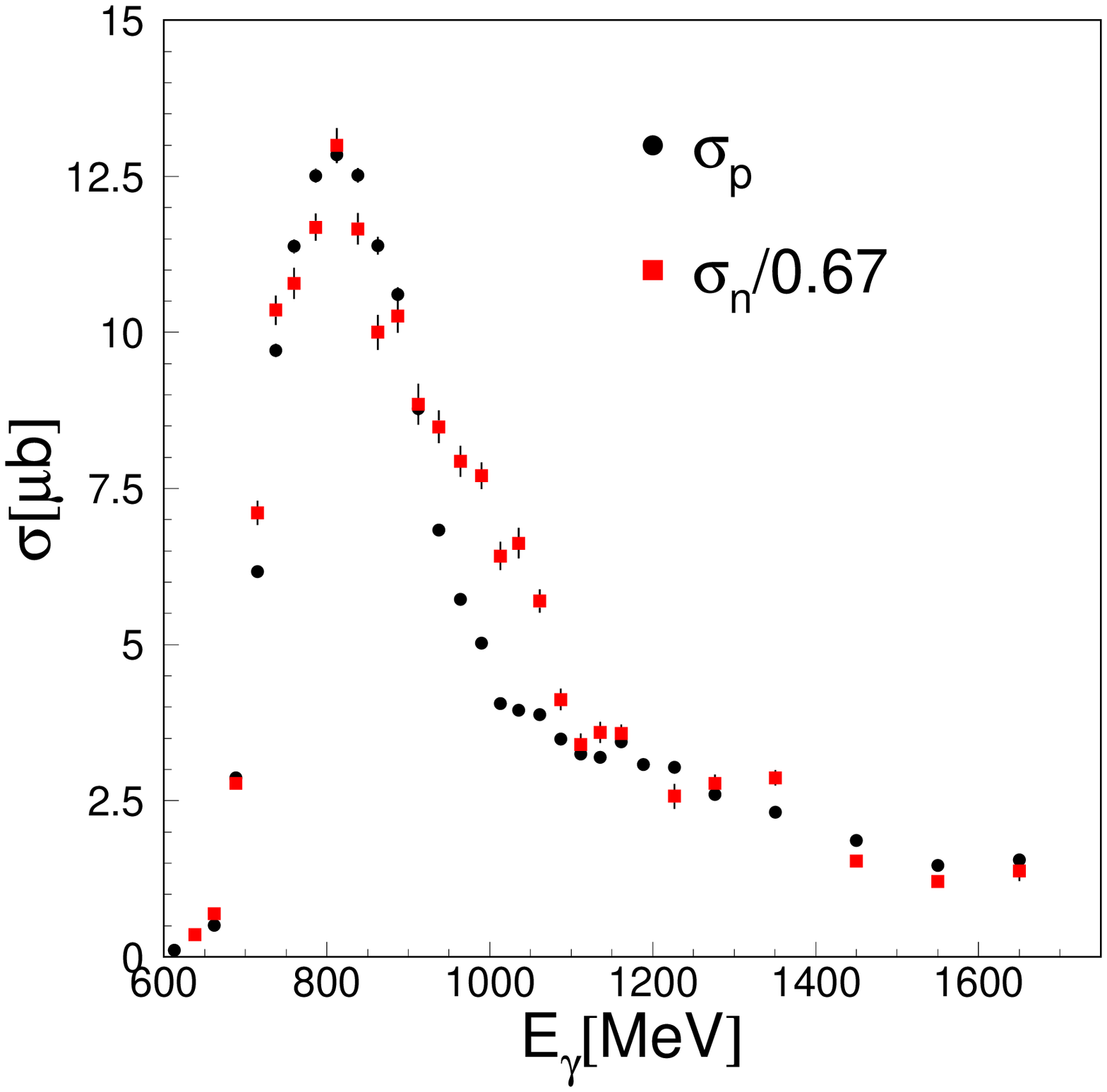,height=3.6cm}
\psfig{file=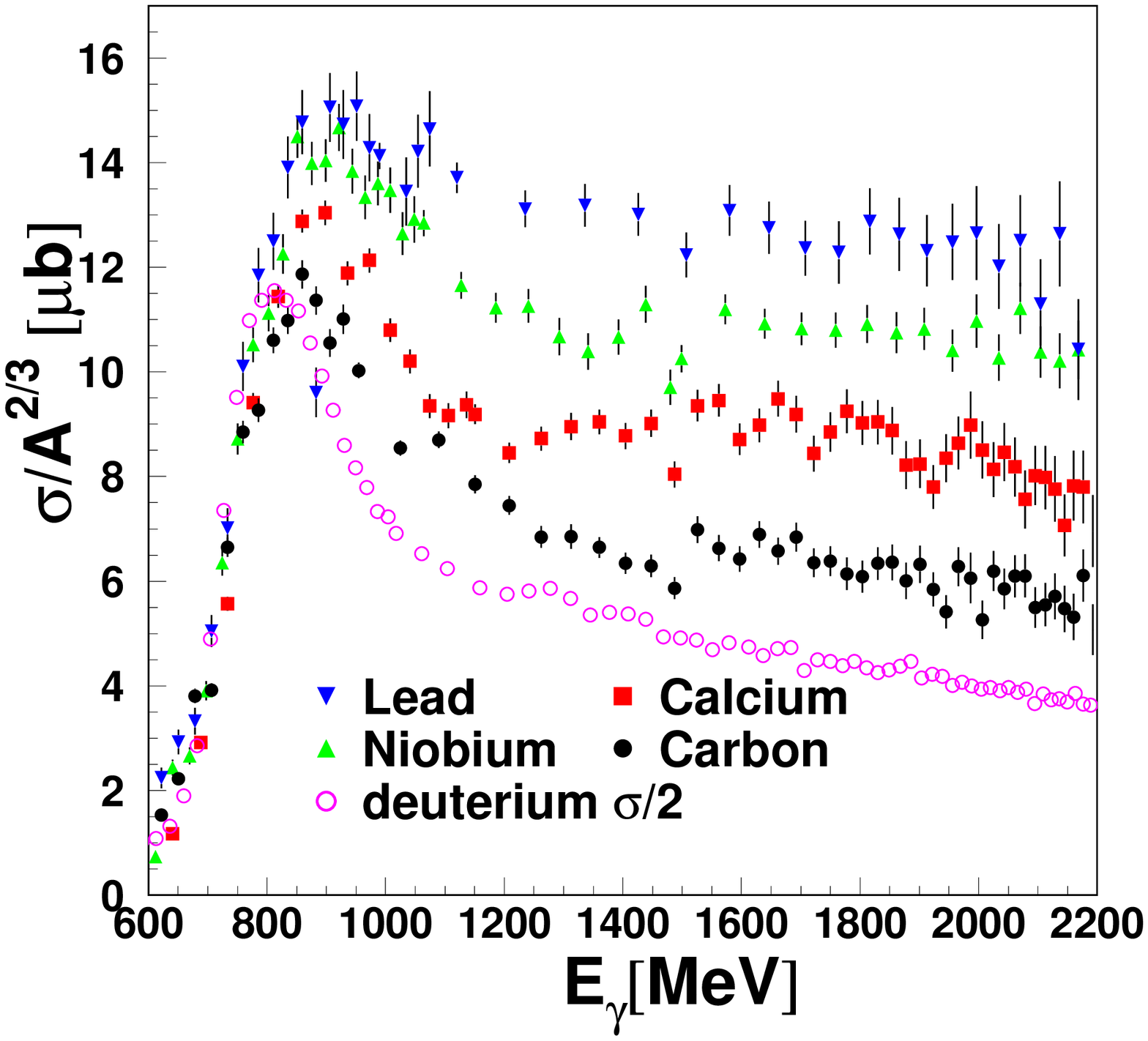,height=3.6cm}
\psfig{file=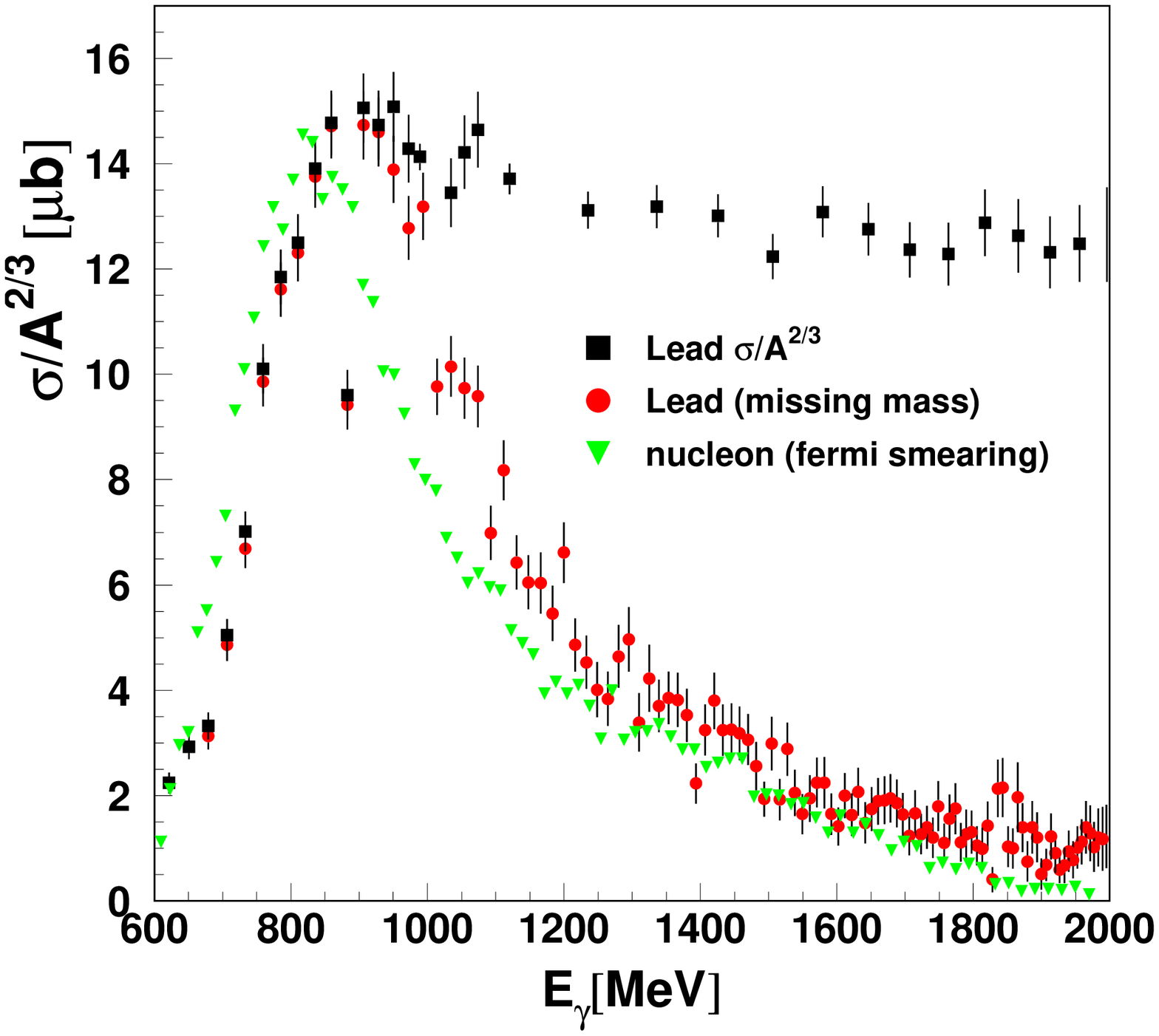,height=3.6cm}}
\vspace*{8pt}
\caption{Left: preliminary result for the total $\gamma p\rightarrow p\eta$ 
and $\gamma n\rightarrow n\eta$ cross sections measured off quasifree nucleons 
from the deuteron.
Center: inclusive $\gamma A\rightarrow \eta X$ cross sections for the deuteron,
carbon, calcium, niobium and lead. Right: Inclusive cross section and single
$\eta$ production cross sections for lead compared to the Fermi smeared average 
nucleon cross section.
\label{fig:eta}}
\end{figure}

Preliminary results for $\eta$ photoproduction off the deuteron and off nuclei 
from the Crystal Barrel/TAPS experiment at the Bonn ELSA accelerator are
summarized in Fig. \ref{fig:eta}. The measurement off quasifree nucleons from the
deuteron (coincident detection of recoil nucleons) shows that
indeed the elementary neutron cross section has a bump-like structure 
(see Fig.~\ref{fig:eta}, left hand side) corresponding to a higher 
lying nucleon resonance. As already reported in Ref.
\refcite{Roebig_96}, the inclusive nuclear cross sections scale like $A^{2/3}$ 
for incident photon energies below 800 MeV. They behave differently 
at higher incident photon energies (see Fig.~\ref{fig:eta}, center) where 
$\eta\pi$ final states and secondary production mechanisms 
(e.g. $\gamma N\rightarrow N\pi$, $\pi N\rightarrow \eta N$) 
contribute and obscure the S$_{11}$
line shape. This contributions can be almost completely suppressed by cuts on 
the reaction kinematics. After these cuts, single $\eta$ photoproduction off 
heavy nuclei like lead (see Fig.~\ref{fig:eta}, right hand side) becomes very 
similar to the Fermi smeared average nucleon cross section. The small 
discrepancy is at least partly due to inefficiencies of the kinematical cuts. 
   
\begin{figure}[hbt]
\centerline{\psfig{file=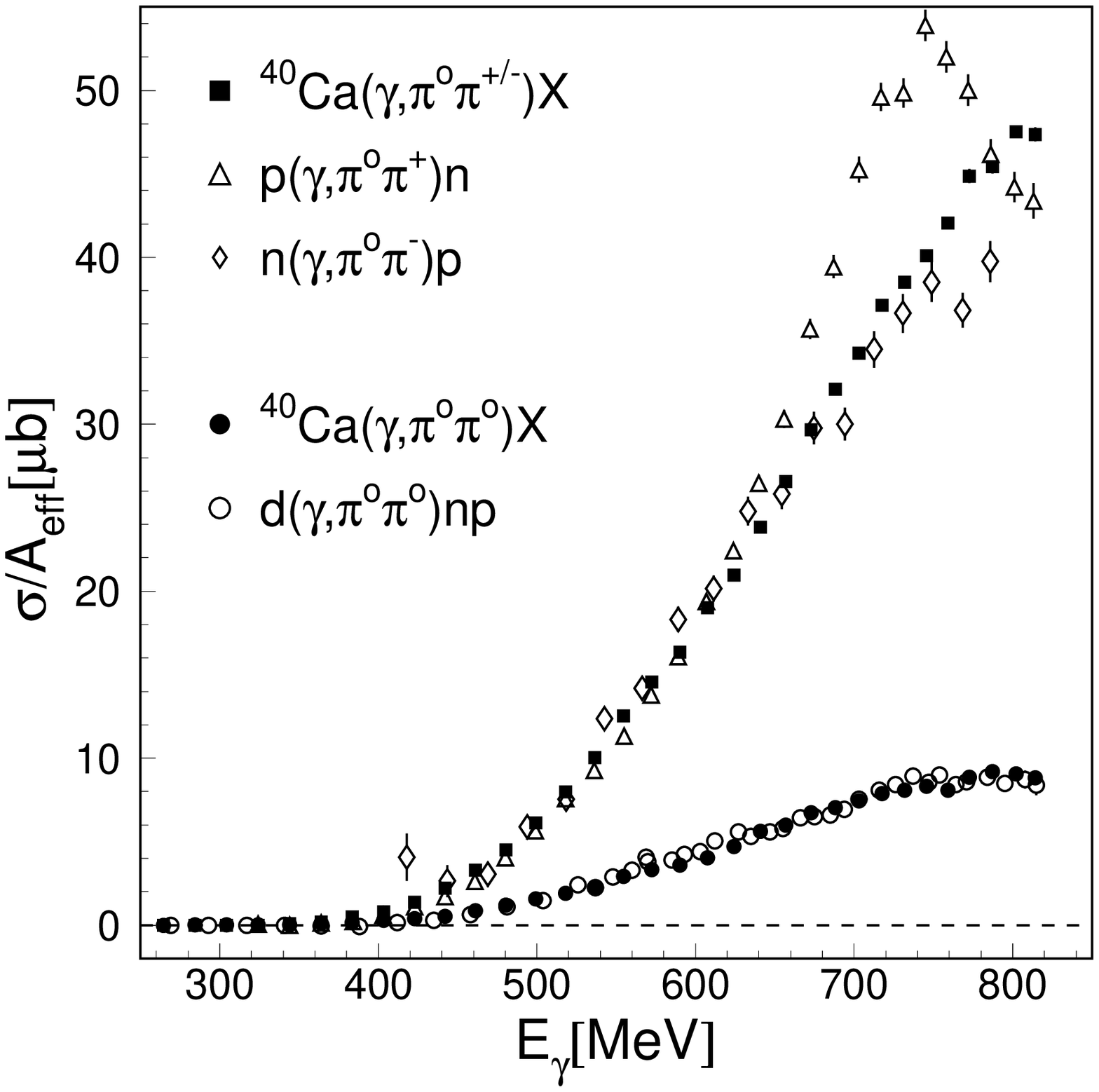,height=3.5cm}
\psfig{file=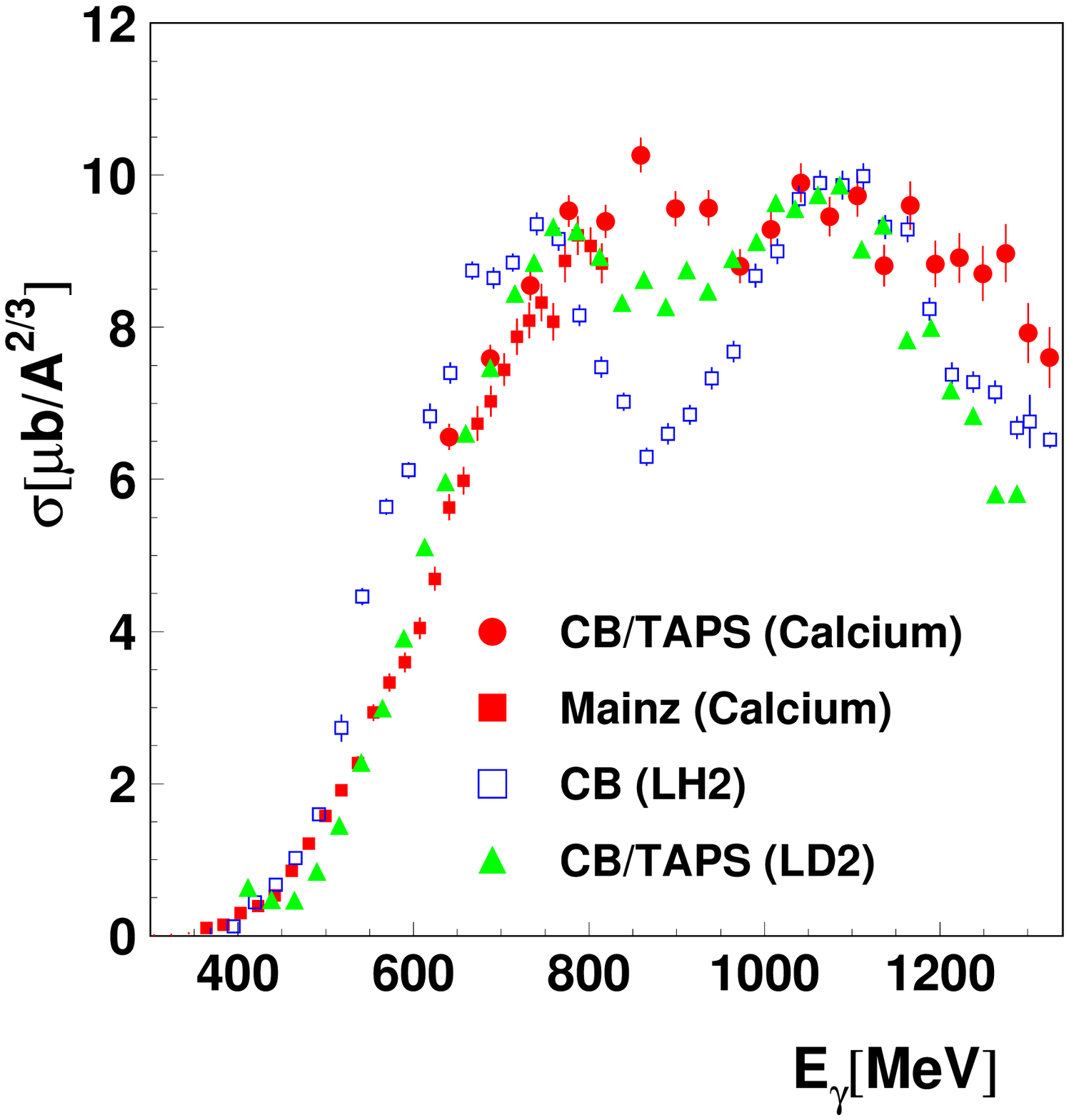,height=3.5cm}}
\vspace*{8pt}
\caption{Left hand side: Total cross sections for $\pi^o\pi^o$ and 
$\pi^o\pi^{\pm}$ photoproduction off $^{40}$Ca compared to 
$d(\gamma ,\pi^o\pi^o)np$ \protect\cite{Kleber_00} respectively 
to $p(\gamma ,\pi^o\pi^+)n$ \protect\cite{Langgaertner_01} and 
$n(\gamma ,\pi^o\pi^-)p$ \protect\cite{Zabrodin_97}. Cross sections normalized to 
$A_{eff}$ = $A^{2/3}$ for calcium and $A_{eff}=A$ for $A=1,2$.
Right hand side: total cross sections for double $\pi^o$ off proton, deuteron
and $^{40}$Ca (preliminary). 
\label{fig:2pspf}}
\end{figure}

Preliminary results for double pion photoproduction are summarized in
Fig.~\ref{fig:2pspf}. Results for double $\pi^o$ and $\pi^o\pi^{\pm}$ have been
obtained\cite{Bloch_06} for incident photon energies up to 800 MeV (see
Fig.~\ref{fig:2pspf}, left hand side). In case of double $\pi^o$, the
excitation functions off the deuteron and off $^{40}$Ca have exactly the same 
shape so that there is no indication for an in-medium modification of the 
D$_{13}$ at energies below its peak position. 
However, the model\cite{Post_04}  
(see Fig.~\ref{fig:selfen}) predicts the main effect for the high energy tail of
the resonance. This part is covered by the preliminary data from the Crystal
Barrel/TAPS experiment (see Fig.~\ref{fig:2pspf}, right hand side), which indeed
shows a difference between deuteron and nuclear data, in particular in the
'dip'-region between the second and third resonance bump. 

Even more interesting than the double $\pi^o$ channel would be the 
$\pi^o\pi^{\pm}$ final state. This is so, because as shown in 
Ref.~\refcite{Lehr_01} the signal for in-medium modifications of resonances is
strongly diluted by the averaging over the nuclear volume for decay channels
that are not directly responsible for the modification. The main effect for
the D$_{13}$ is connected with the coupling to the $N\rho$ channel, and 
only the mixed charge double pion decay can proceed via an intermediate $\rho$ 
(the $\rho^o\rightarrow\pi^o\pi^o$ decay is forbidden). Unfortunately, so far only
data up to 800 MeV are available (see Fig.~\ref{fig:2pspf}, left hand side).
However, even in this region there is already a significant difference in the
excitation functions off the free proton and off $^{40}$Ca. The peak-like 
structure for the proton is suppressed in the nuclear data, but the same 
is true for the quasifree neutron cross section extracted from deuteron data
\cite{Zabrodin_97}. Data over a wider energy range and a comparison of quasifree
data off the bound proton to the bound neutron will be necessary to clarify this
issue.

\section*{Acknowledgments}

The discussed results have been obtained by the TAPS-collaboration at MAMI and
the Crystal Barrel/TAPS collaboration at ELSA. They are part of the theses
works of D. Trnka (U. Giessen) and F. Bloch, I. Jaegle, and Th. Mertens 
(U. Basel). This work was supported by Schweizerischer Nationalfonds.

\end{document}